\documentclass[prl,twocolumn,showpacs,preprintnumbers,amsmath,amssymb,superscriptaddress]{revtex4}

\usepackage{graphicx}
\usepackage{dcolumn}
\usepackage{bm}
\usepackage{bbm}
\usepackage{psfrag}
\usepackage{multirow,amssymb,amsbsy,amsmath}
\usepackage{stmaryrd}

\newcommand{\E}{\textup{e}}
\newcommand{\D}{\textup{d}}
\newcommand*{\Haver}[1]{\mathopen{\llbracket} #1 \mathclose{\rrbracket}}
\newcommand{\vek}[1]{\bm{#1}}
\newcommand{\dd}{\text{d}}
\newcommand{\dod}[2]{\frac{\dd #1}{\dd #2}}
\newcommand{\dodtxt}[2]{\dd #1 / \dd #2}

\newcommand{\pop}[2]{\frac{\partial #1}{\partial #2}}

\newcommand{\ddim}{\udelta\kern0.1em}

\newcommand{\beikonst}[2]{\left( #1 \right)_{\kern-0.2em #2}}

\newcommand{\trtxt}[2][]{\text{Tr}_{#1}\{#2\}}
\newcommand*{\bra}[1]{\mathopen{\langle}#1\mathclose{|}}
\newcommand*{\ket}[1]{\mathopen{|}#1\mathclose{\rangle}}

\newcommand{\comutxt}[2]{[#1,#2]}

\begin{document}
\preprint{APS/123-QED}

%
%
\title{Fourier's Law from Schr\"odinger Dynamics}

\author{Mathias Michel}
\affiliation{Institute of Theoretical Physics I, University of Stuttgart, %
             Pfaffenwaldring 57, 70550 Stuttgart, Germany}%
\email{mathias@theo1.physik.uni-stuttgart.de}
\author{G\"unter Mahler}
\affiliation{Institute of Theoretical Physics I, University of Stuttgart, %
             Pfaffenwaldring 57, 70550 Stuttgart, Germany}%
\author{Jochen Gemmer}%
\affiliation{Physics Department, University of Osnabr\"uck, %
             Barbarastr.\ 7, 49069 Osnabr\"uck, Germany}%

\received{15 March 2005}%

\begin{abstract}
We consider a class of one-dimensional chains of weakly coupled many level systems. 
We present a theory which predicts energy diffusion within these chains for almost all initial states, if some concrete conditions on their Hamiltonians are met. 
By numerically solving the time dependent Schr\"odinger equation, we verify this prediction. 
Close to equilibrium we analyze this behavior in terms of heat conduction and compute the respective coefficient directly from the theory. 
\end{abstract}

\pacs{05.60.Gg, 44.10.+i, 66.70.+f}
\maketitle

%
%

Almost two hundred years ago Fourier conjectured that temperature (or as we know today: energy) tends to diffuse through solids once close enough to equilibrium. 
His results may be stated in the following form
\begin{equation}
  \label{eq:1}
  \pop{}{t}u[T(\vek{x},t)]
  = c(T)\, \pop{}{t}T(\vek{x},t)
  = \vek{\nabla} \cdot (\kappa \vek{\nabla} T)\;,
\end{equation}
where $u[T(\vek{x},t)]$ is the energy density at point $\vek{x}$ and time $t$, $T$ the temperature, $c(T)=\pop{u}{T}$ the specific heat capacity and $\kappa$ the thermal conductivity. 
Despite the ubiquitous occurrence of this phenomenon, its explanation on the basis of some reversible microscopic dynamics remains a serious problem \cite{Garrido2001,Lepri2003}.

One approach to this subject is the Peierls-Boltzmann theory \cite{Peierls1955}. 
To explain the emergence of energy diffusion through an isolator based on quantum theory, Peierls essentially proposed a modified Boltzmann equation. 
He replaced classical particles by quantized quasiparticles such as phonons and assumed statistical transition rates taken from Fermi's Golden Rule for the collision term. 
This concept faces conceptual shortcomings; though quantized normal modes are treated as classical particles, i.e., as being always well localized in configuration as well as momentum space. 
Furthermore, in order to exploit Fermi's Golden Rule in a classical picture, the complete actual quantum state of a phonon mode is discarded and only the mean occupation number is kept which is then treated as classical number of particles. 
Due to the neglect of any phases, this is called the random phase approximation and, as Peierls himself points out, so far lacks concrete justification.

Another concept adressing the occurrence of regular heat conduction is the Green-Kubo formula. Derived on the basis of linear response theory it has originally been formulated for electrical transport \cite{Kubo1957,Kubo1991}. 
In this context the current is viewed as the response to a perturbative electrical potential which can be expressed as a part of the system Hamiltonian. 
But eventually the Green-Kubo formula boils down to a current-current auto-correlation, thus it can {\it ad hoc} be transfered to heat transport simply by replacing the electrical current by a heat current \cite{Luttinger1964}. 
However, the justification of this replacement remains an open problem since there is no way of expressing a temperature gradient in terms of an addend to the Hamiltonian [remarkably enough, Kubo himself comments on that replacement in a rather critical way \cite{Kubo1991}]. 
Despite these unsolved questions, it has become a widely employed technique \cite{Zotos1997,Heidrich2003,Kluemper2002}. 

To overcome this problem such Kubo-scenarios have recently been transfered from Hilbert- to Liouville space, where temperature gradients may be formulated in terms of operators \cite{Michel2005}. 
The method reveals normal heat transport \cite{Saito2002,Michel2003} in very small quantum systems. 
However, it is numerically challenging especially for larger systems. 

\begin{figure}
  \centering 
  \includegraphics[width=7cm]{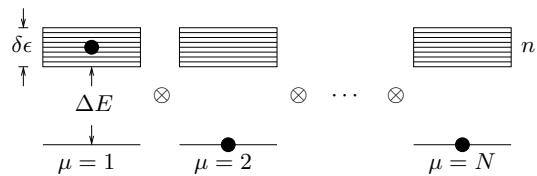}
  \caption{Heat conduction model: $N$ coupled subunits with ground level and band of $n$ equally distributed levels ($\Delta E =1$). Black dots refer to used initial states.}
\label{fig:1}  
\end{figure}

In this letter we introduce yet another approach to heat transport within quantum systems, based on the Hilbert space Average Method (HAM) \cite{Gemmer2003,Gemmer2004,Gemmer2005}. 
To demonstrate the emergence of heat diffusion directly from first principles (Schr\"odinger equation) we apply HAM to an appropriate ``modular design model'', a weakly coupled chain of identical subsystems (cf.\ Fig.~\ref{fig:1}). 
HAM predicts that, if the model meets certain criteria (cf.\ (\ref{eq:11})) the local energy current between two adjacent subunits simply depends linearly on the difference of their inner energies. 
(For spatially small subunits: on the local gradient of the energy.)  
This diffusive behavior is, slightly reformulated [applying  the continuity equation for the energy density $\pop{u}{t}=\vek{\nabla}\cdot\vek{j}$ to (\ref{eq:1})], exactly what Fourier's law states
\begin{equation}
  \label{eq:2}
  \vek{j} 
  = - \kappa \vek{\nabla} T[u(\vek{x},t)]
  = - \kappa \pop{T}{u} \vek{\nabla} u
  = - \frac{\kappa}{c}\vek{\nabla} u(\vek{x},t)\;,  
\end{equation}
where $\vek{j}$ is the energy current density.
To verify HAM we test its predictions for small systems (up to ten subunits) by solving the corresponding time dependent Schr\"odinger equation numerically. 
HAM, however, allows for a distinction between diffusive and non-diffusive behavior and produces the diffusion constant (cf.\ (\ref{eq:10}) directly for arbitrarily large systems.

The class of systems we are going to analyze is depicted in Fig.~\ref{fig:1}, with the Hamiltonian  
\begin{equation}
  \label{eq:3}
  \hat{H} = \sum_{\mu=1}^{N} \hat{H}_{\text{loc}}(\mu) 
          + \lambda\sum_{\mu=1}^{N-1} \hat{V}(\mu,\mu+1) \;.
\end{equation}
Here $N$ identical subunits are assumed to have a non-degenerate ground state and a band of $n$ exited states each, equally distributed over some band width $\delta\epsilon$ such that the band width is small compared to the local energy gap $\Delta E$ (see Fig.~\ref{fig:1}, $\delta\epsilon\ll\Delta E$, $\delta\epsilon$ in units of $\Delta E$).
These subunits are coupled by an energy exchanging next neighbor interaction $\hat{V}(\mu,\mu+1)$, chosen to be a (normalized) random hermitian matrix allowing for any possible transition such as to avoid any bias. 
(Our results will turn out to be independent of the exact form of the matrix.)
We choose the next neighbor coupling to be weak compared to the local gap ($\lambda\ll\Delta E$, $\lambda$ in units of $\Delta E$). 
This way the full energy is approximately given by the sum of the local energies and those are approximately given by $\bra{\psi(t)}\hat{H}_{\text{loc}}(\mu)\ket{\psi(t)} = \Delta E P_{\mu}$ where $P_{\mu}$ is the probability to find the $\mu$'th subsystem in its excited state.
This clean partition of the Hamiltonian into a strong local and a weak interaction part allows for a unique definition of both, a local energy (respectively temperature) as well as a local current.
Again, this oversimplified model is primarily meant to demonstrate the possible direct emergence of diffusive behavior from Schr\"odinger dynamics.
(For the impact of our results on more realistic systems see end of this letter.)
\begin{figure}
  \centering 
  \includegraphics[width=6cm]{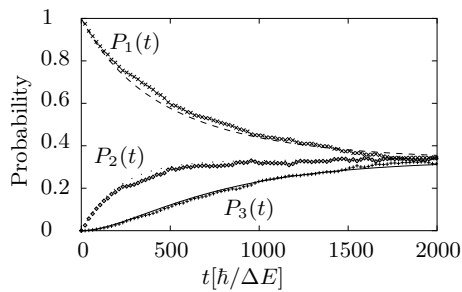}
  \caption{Probability to find the excitation in the $\mu=1,2,3$ system. Comparison of the HAM prediction (lines) and the exact Schr\"odinger solution (dots). ($N=3$, $n=500$, $\lambda=5\cdot 10^{-5}$, $\delta\epsilon=0.05$)} 
\label{fig:2}
\end{figure}

\begin{figure}
  \centering 
  \includegraphics[width=6cm]{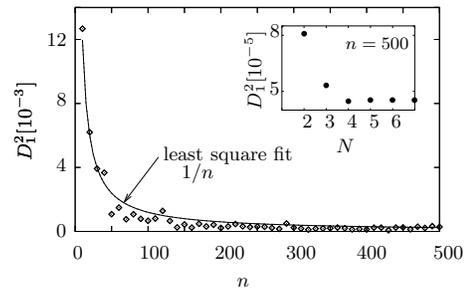}
  \caption{Deviation of HAM from the exact solution ($D^2_1$ least squares): dependence of $D^2_1$ on $n$ for $N=3$. Inset: dependence of $D^2_1$ on $N$ for $n=500$.}
\label{fig:3}  
\end{figure}

The Hilbert space average method (HAM) is in essence a technique to produce a guess for the value of some quantity defined as a function of $\ket{\psi}$ if $\ket{\psi}$ itself is not known in full detail, only some features of it. 
Here it is used to produce a guess for some expectation value $\bra{\psi}\hat{A}\ket{\psi}$ if the only information about $\ket{\psi}$ is the given set of expectation values $\bra{\psi}\hat{P}_{\mu}\ket{\psi}=P_{\mu}$. 
(Thus, $\hat{P}_{\mu}$ is the  projector projecting out the subspace which corresponds to the $\mu$'th subsystem occupying its excitation band, rather than its ground state.)
Such a statement naturally has to be a guess, since there are in general many different $\ket{\psi}$ that are in accord with the given set of $P_{\mu}$ but produce possibly different values for $\bra{\psi}\hat{A}\ket{\psi}$. 
The key question for the reliability of this guess, thus, is whether the distribution of the $\bra{\psi}\hat{A}\ket{\psi}$'s produced by the respective set of $\ket{\psi}$'s is broad or whether almost all those $\ket{\psi}$'s yield $\bra{\psi}\hat{A}\ket{\psi}$'s that are approximately equal. 
It can be shown that if the spectral width of $\hat{A}$ is not too large and  $\hat{A}$ is high-dimensional almost all individual $\ket{\psi}$ yield an expectation value close to the mean of the distribution of the $\bra{\psi}\hat{A}\ket{\psi}$'s and thus the HAM guess will be reliable (see Fig.~\ref{fig:3} and furthermore \cite{Gemmer2004}). 
To find that mean one has to average with respect to the  $\ket{\psi}$'s. 
This is called a Hilbert space average $A$ and denoted as 
\begin{equation}
  \label{eq:4}
  A=\trtxt{\hat{A}\hat{\alpha}}  
  \quad \text{with}\quad 
  \hat{\alpha}
  :=\Haver{\ket{\psi}\bra{\psi}}_{\{\bra{\psi}\hat{P}_{\mu}\ket{\psi}=P_{\mu}\}}\;.
\end{equation}
This expression stands for the average of $\bra{\psi}\hat{A}\ket{\psi}=\trtxt{\hat{A}\ket{\psi}\bra{\psi}}$ over all $\ket{\psi}$ that feature $\bra{\psi}\hat{P}_{\mu}\ket{\psi}=P_{\mu}$ but are uniformly distributed otherwise. 
Uniformly distributed means invariant with respect to all unitary transformations $\hat{U}$ that leave $P_{\mu}$ unchanged, i.e., $\bra{\psi}\hat{U}^{\dagger}\hat{P}_{\mu}\hat{U}\ket{\psi}=P_{\mu}$.
Thus  the $\hat{U}$'s are specified by $\comutxt{\hat{U}}{\hat{P}_{\mu}}=0$.
How is $\hat{\alpha}$ to be computed? 
Any such $\hat{U}$ has to leave $\hat{\alpha}$ invariant, i.e., $\comutxt{\hat{U}}{\hat{\alpha}}=0$. 
Furthermore, $\hat{\alpha}$ has to obey $\trtxt{\hat{\alpha}\hat{P}_{\mu}}=P_{\mu}$. 
Those conditions uniquely determine $\hat{\alpha}$ as
\begin{equation}
  \label{eq:5}
  \hat{\alpha}=\sum_{\mu} \frac{P_{\mu}}{n} \,\hat{P}_{\mu}\;,
\end{equation}
i.e., an expansion in terms of the projectors $\hat{P}_{\mu}$ themselves.

How can this be exploited to describe the dynamics of our system? 
Consider the full system's pure state at some time $t$, $\ket{\psi(t)}$. 
Let $\hat{D}(\tau)$ be a time evolution operator describing the evolution of the system for a short time, i.e., $\ket{\psi(t+\tau)}=\hat{D}(\tau)\ket{\psi(t)}$. 
This allows for the computation of the probabilities $P_{\mu}$ at time $t+\tau$ finding the $\mu$'th system somewhere in its excited band
\begin{equation}
  \label{eq:6}
  P_{\mu}(t+\tau)=\trtxt{\hat{D}(\tau)\,\ket{\psi(t)}\bra{\psi(t)}\,
                         \hat{D}^{\dagger}(\tau)\,\hat{P}_{\mu}}\;.
\end{equation}
Assume that rather than $\ket{\psi(t)}$ itself only the set of expectation values $P_{\mu} $ is known. 
The application of HAM produces a guess for the $P_{\mu} (t+\tau)$ based on the  $P_{\mu}(t)$ through replacing $\ket{\psi(t)}\bra{\psi(t)}$ by   the above $\hat{\alpha}$ 
\begin{equation}
  \label{eq:7}
  P_{\mu}(t+\tau)
  \approx 
  \sum_{\nu}\frac{P_{\nu}}{n}\,
  \trtxt{\hat{D}(\tau)\,\hat{P_{\nu}}\,\hat{D}^{\dagger}(\tau)\,\hat{P_{\mu}}}
  \;.
\end{equation}
In the interaction picture the dynamics of the full system are only controlled by the interaction $\hat{V}(t)$. 
The time evolution is generated by the corresponding Dyson series. 
Thus, assuming weak interactions (\ref{eq:7}) may be evaluated to second order with respect to the interaction strength using an appropriately truncated Dyson series for $\hat{D}(\tau)$. 
This yields after extensive but rather straight forward calculations 
\begin{align}
  \label{eq:8} 
  P_{\mu}(t+\tau)
  &\approx f(\tau)\,[P_{\mu-1}(t)+P_{\mu+1}(t)-2P_{\mu}(t)]\;,\\
  \label{eq:9}
  f(\tau)
  &:= \frac{2}{n}
      \int_0^{\tau}\int_0^{\tau'}
      \trtxt{\hat{V}(\tau'')\,\hat{V}(0)}
      \,\D\tau''\,\D\tau'\;.
\end{align}
The above integrand  is essentially  the same environmental correlation function that appears in the memory kernels of standard projection operator techniques.
Those correlation functions typically feature some decay time $\tau_c$ after which they vanish. 
Thus, i.e, integrating them twice yields functions that increase linear in time after $\tau_c$. 
Hence, for $\tau>\tau_c$ one simply gets $f(\tau)=\gamma\tau$, where $\gamma$ has to be computed from (\ref{eq:9}) but typically corresponds to a transition rate as obtained from Fermi's Golden Rule, i.e.,
\begin{equation}
  \label{eq:10}
  \gamma = \frac{2\pi\lambda^2 n}{\hbar\delta\epsilon}\;.
\end{equation} 
Assuming that the decay times of the correlation functions are small one can transform the iteration scheme (\ref{eq:8}) into a set of differential equations
\begin{align}
  \label{eq:4a}
  \dod{P_1}{t} &= - \gamma(P_1 - P_2)\;, \\
  \label{eq:4b}
  \dod{P_{\mu}}{t} &= - \gamma (2 P_{\mu} - P_{\mu-1} - P_{\mu+1})\;,\\ 
  \label{eq:4c}
  \dod{P_N}{t} &= - \gamma (P_N - P_{N-1})\;.
\end{align}
Applying a discrete version of the continuity equation, $\dodtxt{P_{\mu}}{t}= J_{\mu-1}- J_{\mu}$, to (\ref{eq:4a}-\ref{eq:4c}) yields $J_{\mu}:=- \gamma(P_{\mu+1}-P_{\mu})$. 
This reduces to (\ref{eq:2}), i.e., Fourier's law, with $\gamma=\kappa/c$.

The above scheme, however, only applies if the dynamics of the system are reasonably well described by a Dyson series truncated at second order also for times $\tau$ larger than $\tau_c$. 
This and similar arguments yield the following necessary conditions for the above described occurrence of diffusive transport
\begin{equation}
  \label{eq:11}
  2 \lambda\frac{n}{\delta\epsilon} \geq 1, 
  \quad \lambda^2\frac{n}{\delta\epsilon^2} \ll 1\;.
\end{equation}
We checked by numerically integrating the Schr\"odinger equation (see below) that no diffusive transport as described by (\ref{eq:4a}-\ref{eq:4c}) results if those criteria are violated. 
(For a more detailed description of HAM and its various implications, see \cite{Gemmer2005,Gemmer2004}.) 

To analyze validity and performance of HAM we compare its results  with data from a direct numerical integration of the Schr\"odinger equation. 
This is, of course, only possible for systems small enough to allow for the latter.  
Hereby we restrict ourselves to initial states with only one subsystem in the exited band (all others in the ground level, black dots in Fig.~\ref{fig:1}). 
Finding an effective Hamiltonian for the one-excitation subspace we are able to solve the Schr\"odinger equation for up to $N=10$ subsystems, $n=500$ levels each.
Firstly restricting to $N=3$, the numerical results together with the HAM predictions for an initial state with $P_1=1$, $P_2=P_3=0$ are shown in Fig.~\ref{fig:2} ($N=3$, $n=500$, $\delta\epsilon = 0.05$, $\lambda=5\cdot10^{-5}$). 
There is a reasonably good agreement. 
To investigate the accuracy of the HAM for, e.g., $P_1(t)$ we introduce $D^2_1$, being the time-averaged quadratic deviation of the exact (Schr\"odinger) result for $P_1(t)$ from the HAM prediction
\begin{equation}  
  \label{eq:12}
  D^2_1=\frac{\gamma}{5}
  \int_0^{\frac{5}{\gamma}}(P_1^{\text{HAM}}(t)-P_1^{\text{exact}}(t))^2\D t\;.
\end{equation}
To analyze how big this ``typical deviation'' is, we have computed $D^2_1$ for a $N=3$ system with the first subunit initially in an arbitrary excited state, for different numbers of states $n$ in the bands. 
As shown in  Fig.~\ref{fig:3}, the deviation scales like $1/n$ with the band size, i.e., vanishes in the limit of high dimensional subunits. 
This behavior does not come as a complete surprise since it has theoretically been conjectured and numerically verified in the context of equilibrium fluctuations of non-Markovian systems \cite{Gemmer2004,BorowskiGemmer2003}. 
The inset of Fig.~\ref{fig:3} shows that $D_1^2$ goes down also with increasing number of subunits $N$ but then levels off. 
Altogether HAM appears to be applicable even down to moderately sized systems. 
So far we have restricted ourselves to pure states. 
A drastic further reduction of $D^2$ can be expected for mixed states (which are typical in the context of thermodynamical phenomena), since pure state fluctuations can be expected to cancel partially if added together.

\begin{figure}
  \centering
  \includegraphics[width=6cm]{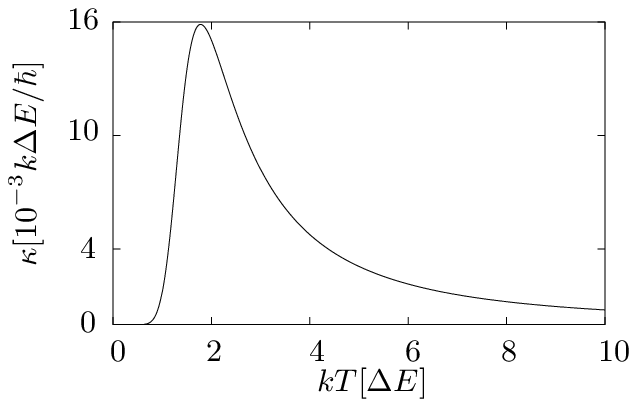}
  \caption{Heat conductivity (\ref{eq:13}) over temperature for a system with $n=500$, $\delta\epsilon = 0.05$ and $\lambda=5\cdot10^{-5}$.}
\label{fig:4}  
\end{figure}
So far we have considered energy diffusion through the system only.
The final state should approach equipartition of energy over all subunits (see Fig.~\ref{fig:2}) -- a thermal equilibrium state (see \cite{Gemmer2004}). 
Close to this equilibrium we expect the system to be in a state where the probability distribution of each subunit is approximately canonical (Gibbs state) but still with a slightly different temperature $T_{\mu}$ for each site (see Fig.~\ref{fig:2} for $t>1000$). 
Specializing in those ``local equilibrium states'' and exploiting the HAM results (\ref{eq:4a}-\ref{eq:4c}) allows for a direct connection of the local energy current between any two adjacent subunits with their temperature gradient $\Delta T=T_1-T_2$ and their mean temperature $T=(T_1+T_2)/2$. 
Since this connection is found to be linear in the temperature gradient one can simply read off the temperature dependent \emph{heat conductivity} (cf.~(\ref{eq:2})) 
\begin{equation}
  \label{eq:13}
  \kappa = \frac{2\pi k\lambda ^2 n^2}{\hbar \delta \epsilon} \;
           \left(\frac{\Delta E}{kT}\right)^2\;
           \frac{\E^\frac{-\Delta E}{kT}}{\Big(1+n\E^\frac{-\Delta E}{kT}\Big)^2}\;,
\end{equation}
as displayed in Fig.~\ref{fig:4}. 
For our (simple) model this is in agreement with $\kappa=\gamma c$ (as
stated after (\ref{eq:4a}-\ref{eq:4c})), if one inserts for $c$ the specific heat of one subunit.
The result of Fig.~\ref{fig:4} is similar to the thermal conductivity of gaped quantum spin chains as calculated by the ``Liouville space method'' \cite{Saito2002}.

What impact do those results for our ``design model'' have on real physical systems? 
For an application of HAM one has to organize the system as a ``net-structure'' of weakly coupled subunits in the  sense of (\ref{eq:11}). 
If this can be established HAM predicts that energy will diffuse from subunit to subunit, irrespective of  whether the coupled subunits form a one- or a more-dimensional net-structure. 
Such a structure may be achieved by coarse graining periodic systems like spin chains, crystals, etc.\ in entities containing many elementary cells and taking the mesoscopic entities as the subunits proper and the effective interactions between the entities as the couplings. 
This way increasing the ``grain size'' will result in higher state densities within the subunits and relatively weaker couplings such that above minimum grain size the criteria (\ref{eq:11}) may eventually be fulfilled.

Of course, the resulting subunits cannot generally be expected to feature the same gaped spectral structure as our design model. 
(Although there are, e.g., spin systems which do so \cite {Saito2002,Schnack2000}.) 
HAM also applies to multi-band subunits. 
Then, for normal transport, the bands and their couplings that mediate the transport should obey the criterion (\ref{eq:11}) individually. 
If this is the case the gaps between the bands are dispensable. 
Only the  band width has to be small compared to the mean band energy. 
Thus, even a continuous spectrum may be spectrally coarse grained to fulfill (\ref{eq:11}). 
This generally yields different energy diffusion constants $\gamma$ for different bands. 
Nevertheless a concrete temperature dependent conductivity $\kappa$ would result for close to equilibrium states. 
Thus HAM should be applicable also to more realistic systems, and work in that direction is under way.

In conclusion we have shown that already a simple modular quantum system can give rise to diffusive energy transport. 
These results are attained based on a new efficient approximation scheme and confirmed by a full Schr\"odinger analysis.

We thank M.\ Hartmann, M.\ Henrich, Ch.\ Kostoglou, H. Michel, H.\ Schmidt, M.\ Stollsteimer and F. Tonner for fruitful discussions. Financial support by the Deutsche Forschungsgesellschaft is gratefully acknowledged.


\end{document}